\documentclass[aps,prd]{revtex4}

\usepackage{graphicx}

\def\lg{{\mathchoice{~\raise.58ex\hbox{$<$}\mkern-14.8mu\lower.52ex\hbox{$>$}~}
                    {~\raise.58ex\hbox{$<$}\mkern-14.8mu\lower.52ex\hbox{$>$}~}
                    {\raise.59ex\hbox{{$\scriptscriptstyle <$}}\mkern-12.8mu%
                     \lower.01ex\hbox{{$\scriptscriptstyle >$}}}   {}   }}
\def\gl{{\mathchoice{~\raise.58ex\hbox{$>$}\mkern-12.8mu\lower.52ex\hbox{$<$}~}
                    {~\raise.58ex\hbox{$>$}\mkern-12.8mu\lower.52ex\hbox{$<$}~}
                    {\raise.62ex\hbox{{$\scriptscriptstyle >$}}\mkern-12.0mu%
                     \lower.05ex\hbox{{$\scriptscriptstyle <$}}}  {}    }}

\newcommand{\be}{\begin{equation}}
\newcommand{\ee}{\end{equation}}
\newcommand{\ba}{\begin{eqnarray}}
\newcommand{\ea}{\end{eqnarray}}
\newcommand{\ban}{\begin{eqnarray*}}
\newcommand{\ean}{\end{eqnarray*}}
\newcommand \nn {\nonumber}
\newcommand{\sla}{\!\!\!/ \,}

\begin{document}

\title{Collective Excitations of Supersymmetric Plasma}

\author{Alina Czajka}

\affiliation{Institute of Physics, Jan Kochanowski University,
25-406 Kielce, Poland}

\author{Stanis\l aw Mr\' owczy\' nski}

\affiliation{Institute of Physics, Jan Kochanowski University,
25-406 Kielce, Poland}
\affiliation{So\l tan Institute for Nuclear Studies, 00-681 Warsaw, Poland}

\date{January 24, 2011}

\begin{abstract}

Collective excitations of ${\cal N} =1$ supersymmetric electromagnetic plasma are studied. 
Since the Keldysh-Schwinger approach is used, not only equilibrium but also non-equilibrium 
plasma, which is assumed to be ultrarelativistic, is under consideration. The dispersion 
equations of photon, photino, electron, and selectron modes are written down and the 
self-energies, which enter the equations, are computed in the Hard Loop Approximation. 
The self-energies are discussed in the context of effective action which is also given. 
The photon modes and electron ones appear to be the same as in the usual ultrarelativistic 
plasma of electrons, positrons and photons. The photino modes coincide with the electron 
ones and the selectron modes are as of a free relativistic massive particle.

\end{abstract}



\maketitle

\section{Introduction}

Supersymmetry is commonly believed to be a symmetry of Nature at a sufficiently high
energy scale. Obviously the symmetry must be broken, as the superpartners of particles,
which constitute the Standard Model, are not seen. Experiments at the Large Hadron
Collider might soon provide evidence of superparticles, but even if this is not the case,
supersymmetry has proved to be a very useful concept of theoretical physics. The
conjectured equivalence -  known as the AdS/CFT duality - of the five-dimensional
gravity in the anti de Sitter geometry and the conformal field theories, see the review
\cite{Aharony:1999ti} and the lecture notes \cite{Klebanov:2000me} as an introduction,
stimulated a great interest in the ${\cal N} =4$ supersymmetric Yang-Mills theory. 
The duality has provided a unique tool to study strongly coupled field theories, as the 
gravitational constant is inversely proportional to the coupling constant  of dual 
conformal field theory and thus some problems of strongly coupled field theories 
can be solved via weakly coupled gravity. Some intriguing results have been obtained 
in this way, see the reviews \cite{Son:2007vk,Janik:2010we},  but relevance of the results 
for non-supersymmetric theories, which are of our actual interest, remains an open issue.
One asks how properties of the supersymmetric quark-gluon plasma governed by 
${\cal N} =4$ SUSY QCD are related to those of the usual quark-gluon plasma experimentally 
studied in relativistic heavy-ion collisions. While such a comparison is, in general, a difficult 
task, some comparative analyses have been done in the domain of weak coupling where 
perturbative methods are applicable \cite{CaronHuot:2006te,CaronHuot:2008uh,Blaizot:2006tk,Chesler:2006gr,Chesler:2009yg}. 
In particular, the paper \cite{Chesler:2009yg} discusses the dispersion relation of quarks
and squarks in equilibrium plasma using the imaginary-time formalism. It is also the 
aim of this paper to compare collective excitations of supersymmetric plasma to those 
of a non-supersymmetric counterpart. However, we study non-equilibrium plasmas where 
the spectrum of excitations is much richer than the equilibrium spectrum. In particular, there 
are unstable modes which dominate the plasma's dynamics. To simplify our  analysis we 
consider the supersymmetric  ${\cal N} =1$ electromagnetic plasma instead  of 
many-body ${\cal N} =4$ SUSY QCD.

There is also another reason for our interest in weakly coupled supersymmetric plasma.
When the plasma is homogeneous but its momentum distribution is anisotropic, there are
instabilities in the gluon sector of quark-gluon plasma or in the photon sector of an 
electromagnetic one, see {\it e.g.} the review \cite{Mrowczynski:2007hb}. Although a general
proof is missing, there seem to be no unstable modes in the fermion sector of quarks and
electrons, respectively, \cite{Mrowczynski:2001az,Schenke:2006fz}. One wonders what happens
in the supersymmetric plasma. Even though the supersymmetry is well known to be broken at a finite
temperature or density, one can still speculate that a rudimentary symmetry induces instability in 
the photino sector when the photon modes are unstable. We test the hypothesis in this paper.

We start our considerations by writing down the lagrangian of supersymmetric ${\cal N} = 1$ 
massless QED. In Sec.~\ref{sec-lagrangian} the general dispersion equations of photons, 
photinos, electrons, and selectrons are written down and the self-energies, which enter the 
equations, are obtained in the subsequent section. The computation is performed within 
the Keldysh-Schwinger approach which allows one to study equilibrium and non-equilibrium 
systems. Various Green's functions of Keldysh-Schwinger formalism are collected in the Appendix. 
Since we are interested in collective modes, the self-energies are found in the long wavelength 
limit using the Hard Loop approach which for equilibrium systems is reviewed in  \cite{Thoma:1995ju,Blaizot:2001nr,Kraemmer:2003gd} while the generalization to anisotropic 
ones is given in \cite{Mrowczynski:2000ed}. The self-energies, which are computed,  are 
also discussed in the context of Hard Loop effective action which was first derived for 
equilibrium plasmas in \cite{Taylor:1990ia,Braaten:1991gm,Frenkel:1991ts}, see also
\cite{Blaizot:1993be,Kelly:1994dh}, and generalized to anisotropic ones in 
\cite{Pisarski:1997cp,Mrowczynski:2004kv}.  Finally, in Sec.~\ref{sec-modes} 
we discuss the collective modes and compare them to  those of ultrarelativistic QED plasma of 
electrons, positrons and photons. We use the natural system of units with $c= \hbar = k_B =1$; 
the signature of the metric tensor is $(+ - - -)$.

\section{${\cal N}=1$ SUSY QED}
\label{sec-lagrangian}

The lagrangian of ${\cal N}=1$ SUSY QED is known, see {\it e.g.} \cite{Binoth:2002xg},
to be
\ba
{\cal L} &=& -\frac{1}{4}F^{\mu \nu} F_{\mu \nu} 
+  i\bar \Psi D\!\sla \Psi
+\frac{i}{2} \bar \Lambda \partial \sla \Lambda
+(D_\mu \phi_L)^*(D^\mu \phi_L) + (D_\mu^* \phi_R)(D^\mu \phi_R^*)
\\ \nn
&& +\sqrt{2} e \big( \bar \Psi P_R \Lambda \phi_L - \bar \Psi P_L \Lambda \phi_R^*
+ \phi_L^* \bar \Lambda P_L \Psi - \phi_R \bar \Lambda P_R \Psi \big)
- \frac{e^2}{2} \big( \phi_L^* \phi_L - \phi_R^* \phi_R \big)^2 ,
\ea
where the strength tensor $F^{\mu \nu}$ is expressed through the electromagnetic
four-potential $A^\mu$ as $F^{\mu \nu} \equiv \partial^\mu A^\nu - \partial^\nu A^\mu$
and the covariant derivative is $D^\mu \equiv \partial^\mu +ie A^\mu$;
$\Lambda$ is the Majorana bispinor photino field, $\Psi$ is the
Dirac bispinor electron field, $\phi_L$ and $\phi_R$ are the scalar left selectron
and right selectron fields; the projectors $P_L$ and $P_R$ are defined in a standard
way $P_L \equiv \frac{1}{2}(1 - \gamma_5)$ and $P_R \equiv \frac{1}{2}(1 + \gamma_5)$.
Since we are interested in ultrarelativistic plasmas, the mass terms are neglected
in the lagrangian. We note that the quark-gluon plasma, as studied in relativistic 
heavy-ion collisions, is ultrarelativistic and quark masses are usually safely ignored.

\section{Dispersion equations}
\label{sec-dis-eqs}

Dispersion equations determine dispersion relations of quasi-particle excitations. Below 
we write down the dispersion equation of quasi-photons, quasi-electrons, quasi-photinos, 
and quasi-selectrons.

\subsection{Photons}

Since the equation of motion of the electromagnetic field $A^{\mu}(k)$  is of the form
\be
\label{eq-motion-A}
\Big[ k^2 g^{\mu \nu} -k^{\mu} k^{\nu} - \Pi^{\mu \nu}(k) \Big]
A_{\nu}(k) = 0 ,
\ee
where $\Pi^{\mu \nu}(k)$ is the retarded polarization tensor and $k\equiv (\omega, {\bf k})$
is the four-momentum, the general photon dispersion equation is
\be
\label{dis-photon-1}
{\rm det}\Big[ k^2 g^{\mu \nu} -k^{\mu} k^{\nu} - \Pi^{\mu \nu}(k) \Big]
 = 0 \;.
\ee
Equivalently, the dispersion relations are given by positions of poles of effective
photon propagator. Because of the transversality of $\Pi^{\mu \nu}$ ($k_\mu \Pi^{\mu \nu}(k) =0$),
which is required by gauge covariance, not all components of $\Pi^{\mu \nu}$ are independent
from each other and consequently the dispersion equation (\ref{dis-photon-1})  can be much
simplified by expressing the polarization tensor through the dielectric tensor $\varepsilon^{ij}(k)$.

\subsection{Electrons} 

The electron field $\psi (k)$ obeys the equation
\be 
\Big[ k\sla  - \Sigma (k)  \Big] \psi (k) =0 , 
\ee
 where $\Sigma (k)$ is the retarded electron self-energy,
and thus the dispersion equation is
\be
\label{dis-electron-1}
 {\rm det}\Big[ k\sla  - \Sigma (k) \Big]  = 0 . 
\ee
Further on we assume that the spinor structure of
$\Sigma(k)$ is 
\be 
\label{structure-electron} 
\Sigma (k) = \gamma^{\mu} \Sigma_{\mu}(k) .
\ee 
Then, substituting the expression (\ref{structure-electron}) into
Eq.~(\ref{dis-electron-1}) and computing the determinant as
explained in Appendix 1 of \cite{Mrowczynski:1992hq}, we get 
\be
\label{dis-electron-2} 
\Big[\big( k^{\mu} - \Sigma^{\mu}(k)
\big) \big(k_{\mu} - \Sigma_{\mu}(k) \big)\Big]^2  = 0 . 
\ee

\subsection{Photinos}

The photino equation of motion is
\be 
\Big[ k\sla  - \tilde \Pi (k)  \Big] \Lambda (k) =0, 
\ee
where $\Lambda$ is the photino Majorana bispinor and $\tilde \Pi$
is the retarded self-energy. The dispersion equation is 
\be
\label{dis-photino-1} {\rm det}\Big[ k\sla  - \tilde \Pi (k)
\Big] = 0 . 
\ee 
Since the expected spinor structure of $\tilde \Pi(k)$ is analogous to
that given by Eq.~(\ref{structure-electron}), the dispersion equation 
coincides with Eq.~(\ref{dis-electron-2}).

\subsection{Selectrons}

The selectron fields $\phi_L (k)$ and $\phi_R (k)$ obey the
Klein-Gordon equation
\be 
\label{dis-eq-selectron}
\Big[ k^2  + \tilde\Sigma_{L,R} (k) \Big] \phi_{L,R} (k) =0 ,
\ee

where $\tilde\Sigma_{L,R} (k)$ is the retarded self-energy of left
or right selectrons. The dispersion equation is
\be 
\label{dis-selectron}
k^2 + \tilde\Sigma_{L,R} (k) = 0 .
 \ee

\section{Self-energies}
\label{sec-self-energies} 

In this section we compute the self-energies which enter the dispersion 
equations (\ref{dis-photon-1}, \ref{dis-electron-1}, \ref{dis-photino-1}, \ref{dis-selectron}).
The plasma is assumed to be homogeneous but the momentum distribution is, in general, 
different from equilibrium one. Therefore, we use the the Keldysh-Schwinger formalism
and the free Green's functions, which are labeled with the indices $+,-, >, <, {\rm sym}$, 
are collected in the Appendix. The computation is performed within the Hard Loop approach, 
see the reviews  \cite{Thoma:1995ju,Blaizot:2001nr,Kraemmer:2003gd}, which was 
generalized to anisotropic systems in \cite{Mrowczynski:2000ed}. The plasma is assumed 
to be ultrarelativistic and thus masses of electrons and selectrons are neglected. We also 
assume that the system is neutral and that the distribution function of electrons ($f_e({\bf p})$) 
equals the distribution function of positrons ($\bar f_e({\bf p})$). Analogous equality is assumed 
for selectrons:  $f_s({\bf p}) = \bar f_s({\bf p})$. The additional assumption is that both left and 
right selectrons are described by the same function $f_s({\bf p})$.

\subsection{Polarization tensor}

\begin{figure}[t]
\centering
\includegraphics*[width=0.7\textwidth]{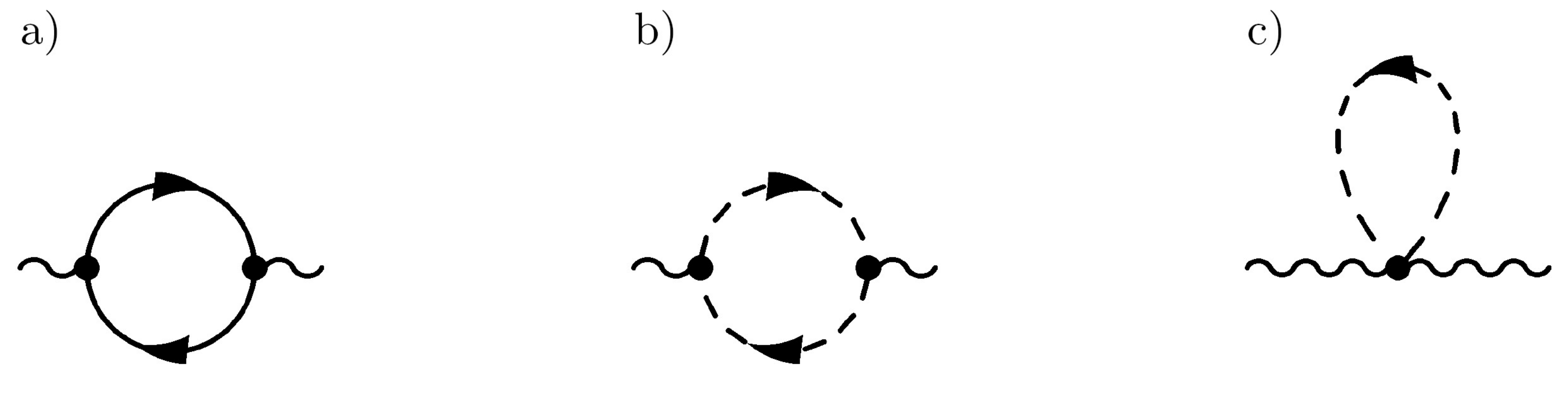}
\caption{Contributions to the photon self-energy. }
\label{fig-photon}
\end{figure}

The polarization tensor $\Pi^{\mu \nu}$ can be defined by means
of the Dyson-Schwinger equation
\be
i{\cal D}^{\mu \nu} (k) = i D^{\mu \nu} (k)
+ i D^{\mu \rho}(k) \, i\Pi_{\rho \sigma}(k)  \, i{\cal D}^{\sigma \nu}(k) ,
\ee
where ${\cal D}^{\mu \nu}$ and $D^{\mu \nu}$ is the interacting and free photon
propagator, respectively. The lowest order contributions to $\Pi^{\mu \nu}$ are given 
by the three diagrams shown in Fig.~\ref{fig-photon}. The solid, wavy and dashed lines 
denote, respectively, electron, photon and selectron fields.

\subsubsection{Electron loop}

Applying the Feynman rules of the Keldysh-Schwinger formalism, which are discussed
in {\it e.g.} Sec.~8 of \cite{Mrowczynski:1992hq}, the contribution to $\Pi_{\mu \nu}^{\lg}$
from the electron loop corresponding to the graph from Fig.~\ref{fig-photon}a is immediately 
written down in the coordinate space as
\be
i {_{(a)}\Pi_{\mu \nu}^{\lg}} (x) = (-1)(-ie)^2
{\rm Tr} [\gamma _\mu iS^{\lg} (x) \gamma _\nu iS^{\gl}(-x)] ,
\ee
where the factor $(-1)$ occurs due to the fermion loop. It gives
\be
 {_{(a)}\Pi_{\mu \nu}^{\lg}} (x) = -i e^2
{\rm Tr} [\gamma _\mu S^{\lg} (x) \gamma _\nu S^{\gl}(-x)] .
\ee
Since
\be
\Pi_{\mu \nu}^{\pm} (x) = \pm \Theta(\pm x_0)
\Big( \Pi_{\mu \nu}^> (x) - \Pi_{\mu \nu}^< (x) \Big),
\;\;\;\;\;\;\;\;\;\;\;\;
S^{\pm} (x) = \pm \Theta(\pm x_0)
\Big(S^> (x) - S^< (x) \Big)
\ee
the retarded polarization tensor $_{(a)}\Pi_{\mu \nu}^+ (x)$ is found as
\be
\label{Pi-x-1}
_{(a)}\Pi_{\mu \nu}^+ (x) = i \frac{e^2}{2}
{\rm Tr}\big[\gamma _\mu S^+(x) \gamma _\nu S^{\rm sym}(-x)
+ \gamma _\mu S^{\rm sym}(x) \gamma _\nu S^-(-x) \big].
\ee
In the momentum space it reads
\be
\label{Pi-k-e-1}
_{(a)}\Pi^{\mu \nu}(k) = i\frac{e^2}{2}
 \int \frac{d^4p}{(2\pi )^4}
{\rm Tr} \big[\gamma^\mu S^+(p+k)\gamma ^\nu S^{\rm sym}(p)
+ \gamma ^\mu S^{\rm sym}(p) \gamma ^\nu S^-(p-k) \big].
\ee
The index $+$ of the polarization tensor from Eq.~(\ref{Pi-k-e-1}) is dropped.
Further on, we will consider only the retarded self-energies and thus
the index $+$ will not be used.

Substituting the functions $S^{\pm}$ (\ref{S-pm}) and $S^{\rm sym}$ (\ref{S-sym}) 
into Eq.~(\ref{Pi-k-e-1}), one finds
\ba
\label{Pi-k-e-3}
_{(a)}\Pi^{\mu \nu}(k) &=&
-\frac{e^2}{4}
 \int \frac{d^3p}{(2\pi )^3} \, \frac{2f_e({\bf p}) -1}{E_p}
\\ \nn
&&\times
{\rm Tr} \bigg[
\bigg(
\frac{ \gamma^\mu (p\sla + k\sla)\gamma ^\nu  p\sla
+ \gamma ^\mu p\sla \gamma ^\nu(p\sla + k\sla)}
{(p+k)^2 + i\, {\rm sgn}\big((p+k)_0\big)0^+}
+  \frac{\gamma ^\mu p\sla \gamma ^\nu(p\sla -k\sla)
+  \gamma^\mu (p\sla - k\sla)\gamma ^\nu  p\sla}
{(p-k)^2 - i\, {\rm sgn}\big((p-k)_0\big)0^+}
\bigg) \bigg],
\ea
where after performing the integration over $p_0$, the momentum ${\bf p}$ 
was changed into  $-{\bf p}$ in the positron contribution. It was also assumed 
that $f_e({\bf p}) = \bar f_e({\bf p}) $. 

Computing the traces of gamma matrices and taking into account that $p^2 =0$, 
one finds
\ba
\label{Pi-k-e-4}
_{(a)}\Pi^{\mu \nu}(k) &=&
- 2e^2  \int \frac{d^3p}{(2\pi )^3} \, \frac{2f_e({\bf p}) -1}{E_p}
\\ \nn
&& \times
\bigg(
\frac{2p^\mu p^\nu  + k^\mu p^\nu + p^\mu k^\nu - g^{\mu \nu} (k \cdot p)  }
{(p+k)^2 + i\, {\rm sgn}\big((p+k)_0\big)0^+}
+ \frac{2p^\mu p^\nu  -  k^\mu p^\nu - p^\mu k^\nu + g^{\mu \nu} (k \cdot p) }
{(p-k)^2 - i\, {\rm sgn}\big((p-k)_0\big)0^+}
\bigg) .
\ea

We are interested in collective modes which occur when the wavelength of
a quasi-particle is much bigger than a characteristic interparticle distance
in the plasma. Thus, we look for the polarization tensor at $k^\mu \ll p^\mu$
which is the condition of the Hard Loop Approximation for anisotropic
systems \cite{Mrowczynski:2000ed,Mrowczynski:2004kv}.  The approximation
is implemented by observing that 
\ban
\frac{1}{(p+k)^2 + i0^+}
+ \frac{1} {(p-k)^2 - i0^+}
= \frac{2k^2}{(k^2)^2 - 4 (k\cdot p)^2 - i {\rm sgn}(k\cdot p) 0^+}
\approx -\frac{1}{2}  \frac{k^2}{(k\cdot p + i 0^+)^2} ,
\\ [2mm]
\frac{1}{(p+k)^2 + i0^+}
- \frac{1} {(p-k)^2 - i0^+}
= \frac{4(k \cdot p)}{(k^2)^2 - 4 (k\cdot p)^2 - i {\rm sgn}(k\cdot p) 0^+}
\approx \frac{k\cdot p}{(k\cdot p + i 0^+)^2}.
\ean
We note that $(p+k)_0 > 0$, $(p-k)_0 > 0$ for $p^\mu \gg k^\mu$.
With the above formulas Eq.~(\ref{Pi-k-e-4}) gives
\ba
\label{Pi-k-e-final}
_{(a)}\Pi^{\mu \nu}(k) &=&
2e^2  \int \frac{d^3p}{(2\pi )^3} \, \frac{2f_e({\bf p})-1}{E_p} \,
\frac{k^2 p^\mu p^\nu  -  \big(k^\mu p^\nu + p^\mu k^\nu 
- g^{\mu \nu} (k \cdot p) \big) (k \cdot p)}
{(k\cdot p + i 0^+)^2} ,
\ea
which is the well-known form of polarization tensor of photons and of gluons 
in ultrarelativistic plasmas, see {\it e.g.} the reviews 
\cite{Mrowczynski:2007hb,Blaizot:2001nr}. As seen, $_{(a)}\Pi^{\mu \nu}(k)$ 
is symmetric ($_{(a)}\Pi^{\mu \nu}(k) = {_{(a)}}\Pi^{\nu \mu}(k)$) and transverse 
($k_\mu {_{(a)}}\Pi^{\mu \nu}(k) = 0$) as required by the gauge invariance.

\subsubsection{Selectron loop}

The contribution to the polarization tensor coming from the selectron loop depicted
in Fig.~\ref{fig-photon}b is given by an appropriately modified Eq.~(\ref{Pi-k-e-1}) that is
\be
\label{Pi-k-s-1}
_{(b)}\Pi^{\mu \nu}(k) = - i\frac{e^2}{2}
 \int \frac{d^4p}{(2\pi )^4}
\big[(2p+k)^\mu (2p+k)^\nu \tilde S^+(p+k) \tilde S^{\rm sym}(p)
+ (2p-k)^\mu (2p-k)^\nu \tilde S^{\rm sym}(p) \tilde S^-(p-k) \big].
\ee
The sign is different than in Eq.~(\ref{Pi-k-e-1}), as we deal here with the
boson not the fermion loop. Substituting the functions $\tilde S^{\pm}$ and
$\tilde S^{\rm sym}$ given by Eqs.~(\ref{Del-pm}, \ref{Del-sym}) into
Eq.~(\ref{Pi-k-s-1}), one finds
\be
\label{Pi-k-s-3}
_{(b)}\Pi^{\mu \nu}(k) = -\frac{e^2}{2}  \int \frac{d^3p}{(2\pi )^3} \,
\frac{2f_s({\bf p})+1}{E_p} \,
\bigg[  \frac{(2p+k)^\mu (2p+k)^\nu}{(p+k)^2 + i\, {\rm sgn}\big((p+k)_0\big)0^+}
+  \frac{(2p-k)^\mu  (2p-k)^\nu}{(p-k)^2 - i\, {\rm sgn}\big((p-k)_0\big)0^+}\bigg],
\ee
where the change ${\bf p} \rightarrow -{\bf p}$ was made in the antiselectron part
and we assumed that $\bar f_s({\bf p}) = f_s({\bf p})$. After adopting the Hard Loop 
Approximation Eq.~(\ref{Pi-k-s-3}) gives
\be
\label{Pi-k-s-4}
_{(b)}\Pi^{\mu \nu}(k) =  e^2  \int \frac{d^3p}{(2\pi )^3} \, \frac{2f_s({\bf p})+1}{E_p} \,
 \frac{k^2 p^\mu p^\nu - (p^\mu k^\nu + k^\mu p^\nu)(k \cdot p)}{(k \cdot p + i0^+)^2}.
\ee

\subsubsection{Selectron tadpole}
The contribution to the polarization tensor coming from the selectron tadpole depicted
in Fig.~\ref{fig-photon}c is
\be
\label{Pi-k-s-t-1}
i {_{(c)}}\Pi^{\mu \nu}(k) = 2i e^2 g^{\mu \nu}
 \int \frac{d^4p}{(2\pi )^4}  i\tilde S^<(p) .
\ee Substituting the function $\tilde S^<$ given by
Eq.~(\ref{Del-<}) into Eq.~(\ref{Pi-k-s-t-1}), one finds \be
\label{Pi-k-s-t-2} _{(c)}\Pi^{\mu \nu}(k) =  e^2 g^{\mu \nu}
 \int \frac{d^3p}{(2\pi )^3}  \,
\frac{2f_s({\bf p})+1}{E_p}  ,
\ee
where the equality $\bar f_s({\bf p}) = f_s({\bf p})$ was assumed.

We get the complete contribution from a single selectron field to the polarization tensor
by summing the contributions from the selectron loop and the selectron tadpole. Thus,
one finds
\be
\label{Pi-k-s-total}
_{(b+c)}\Pi^{\mu \nu}(k) =  e^2  \int \frac{d^3p}{(2\pi )^3} \,
\frac{2f_s({\bf p})+1}{E_p} \,
 \frac{k^2 p^\mu p^\nu - \big(p^\mu k^\nu + k^\mu p^\nu - g^{\mu \nu} (k \cdot p)\big)(k \cdot p)}
{(k \cdot p + i0^+)^2}.
\ee
As seen, it is of exactly the same form as the electron contribution given by
Eq.~(\ref{Pi-k-e-final}) -- it is symmetric and transversal. Actually, the expression 
(\ref{Pi-k-e-final}) is the polarization tensor of scalar QED, which for equilibrium plasma
was discussed in {\it e.g.} \cite{Kraemmer:1994az} using the imaginary-time formalism. 
Since there are two selectron fields in ${\cal N} =1$ SUSY QED, the expression 
(\ref{Pi-k-s-total}) should be multiplied by a factor of 2 to get the complete selectron 
contribution to the polarization tensor.

\subsubsection{Final result}

Combining the electron (\ref{Pi-k-e-final}) and selectron (\ref{Pi-k-s-total}) contributions,
we get the final expression of the polarization tensor
\be
\label{Pi-k-final}
\Pi^{\mu \nu}(k) =  4e^2  \int \frac{d^3p}{(2\pi )^3} \,
\frac{f_e({\bf p})+f_s({\bf p})}{E_p} \,
 \frac{k^2 p^\mu p^\nu - \big(p^\mu k^\nu + k^\mu p^\nu - g^{\mu \nu} (k \cdot p)\big)(k \cdot p)}
{(k \cdot p + i0^+)^2}.
\ee
As seen, $\Pi^{\mu \nu}(k)$ vanishes in the vacuum limit when $f_e , f_s \rightarrow 0$.
This is a nice feature of supersymmetric plasma. In the non-supersymmetric counterpart,
the polarization tensor is given by Eq.~(\ref{Pi-k-e-final}) where the vacuum contribution
diverges and it requires a special treatment. Up to the vacuum contribution, 
the polarization tensor of supersymmetric plasma and of its non-supersymmetric 
counterpart has the same structure. 

\begin{figure}[t]
\centering
\includegraphics*[width=0.45\textwidth]{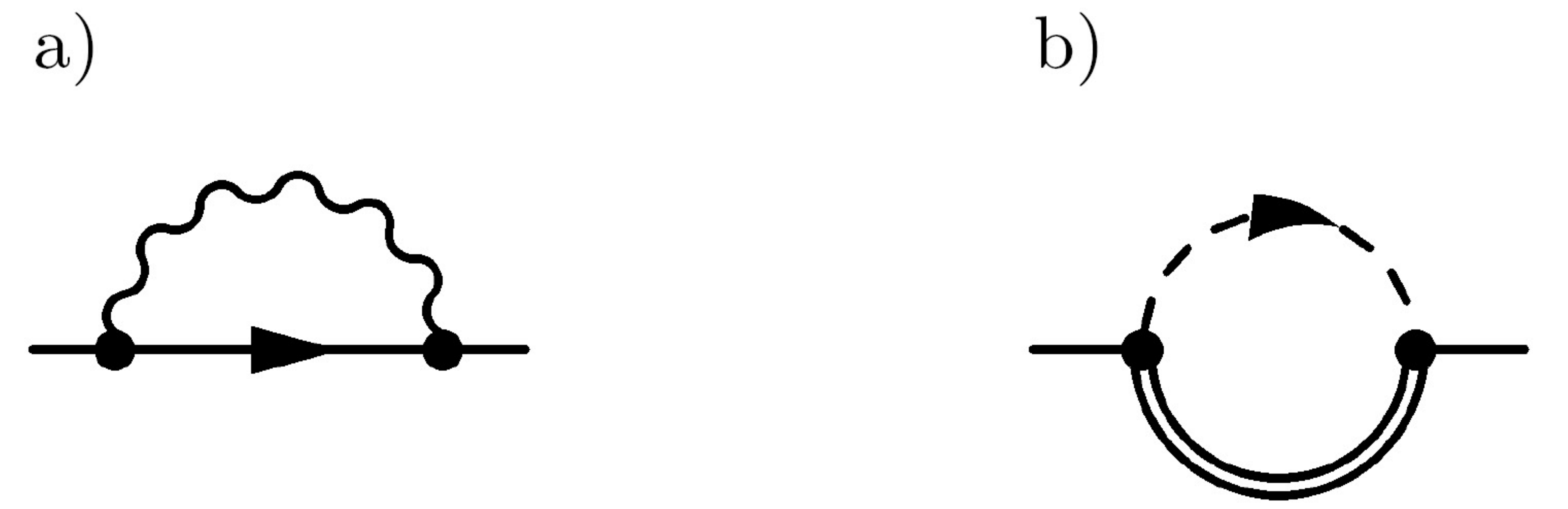}
\caption{Contributions to the electron self-energy. }
\label{fig-electron}
\end{figure}

\subsection{Electron self-energy}

The electron self-energy $\Sigma$ can be defined by means of the Dyson-Schwinger 
equation 
\be 
i{\cal S} (k) = i S (k) + i S(k) \, \big(-i\Sigma (k) \big) \, i{\cal S}(k) , 
\ee 
where ${\cal S}$ and $S$ is the interacting and free propagator, respectively. 
The lowest order contributions to $\Sigma$ are given by two diagrams
shown in Fig.~\ref{fig-electron}. The solid, wavy, dashed and double-solid lines 
denote, respectively, electron, photon, selectron and photino fields.

\subsubsection{Electron-photon loop}

The contribution to the electron self-energy corresponding to the
graph depicted in Fig.~\ref{fig-electron}a is given by an appropriately modified
Eq.~(\ref{Pi-k-e-1}) that is \be -i {_{(a)} \Sigma} (k) = (-ie)^2
\frac{1}{2}
 \int \frac{d^4p}{(2\pi )^4}
\big[\gamma^\mu iS^+(p+k) \gamma^\nu iD^{\rm sym}_{\mu \nu}(p)
+ \gamma^\mu  iS^{\rm sym}(p)  \gamma^\nu iD^-_{\mu \nu}(p-k)
\big],
\ee
which trivially gives
\be
\label{Si-k-a-1}
_{(a)} \Sigma (k) = i e^2 \frac{1}{2}
 \int \frac{d^4p}{(2\pi )^4}
\big[\gamma^\mu S^+(p+k) \gamma^\nu D^{\rm sym}_{\mu \nu}(p)
+ \gamma^\mu  S^{\rm sym} (p) \gamma^\nu D^-_{\mu \nu}(p-k)
\big].
\ee
Substituting the functions $D^{\pm}_{\mu \nu}$, $D^{\rm sym}_{\mu \nu}$
and $S^{\pm}$, $S^{\rm sym}$ given by Eqs.~(\ref{D-pm}, \ref{D-sym}, \ref{S-pm}, \ref{S-sym})
into Eq.~(\ref{Si-k-a-1}), one finds
\ba
\label{Si-k-a-3}
_{(a)} \Sigma (k) &=&  \frac{e^2}{2}
 \int \frac{d^3p}{(2\pi )^3 E_p}
\bigg\{ \bigg[
  \frac{p\sla+k\sla }{(p+k)^2 + i\, {\rm sgn}\big((p+k)_0\big)0^+}
- \frac{p\sla - k\sla}{(p-k)^2 - i\, {\rm sgn}\big((p-k)_0\big)0^+}
\bigg]  \big[2 f_\gamma ({\bf p}) +1\big]
\\ \nn
&& \;\;\;\;\;\;\;\;\;\;\;\;\;\;\;\;\;\;\;\;\;\;\;\;\;-
\bigg[
\frac{p\sla }{(p-k)^2 - i\, {\rm sgn}\big((p-k)_0\big)0^+}
- \frac{p\sla }{(p+k)^2 + i\, {\rm sgn}\big((p+k)_0\big)0^+}
  \bigg] \big[2 f_e({\bf p}) - 1\big]  \bigg\},
\ea
where the change ${\bf p} \rightarrow -{\bf p}$ was made  
in the negative energy terms.  It was also  assumed that 
$f_e({\bf p}) = \bar f_e({\bf p})$. Applying the Hard Loop
Approximation, one obtains
\ba
\label{Si-k-a-4}
_{(a)} \Sigma (k) &=& e^2  \int \frac{d^3p}{(2\pi )^3} \,
 \frac{ f_\gamma ({\bf p}) +  f_e ({\bf p})}{E_p}  \,
\frac{p\sla}{k\cdot p + i 0^+} ,
\ea
which is the well-known form of self-energy  of electrons and of quarks
in ultrarelativistic plasmas, see {\it e.g.} the review \cite{Blaizot:2001nr}.

\subsubsection{Selectron-photino loop}

Since there are two selectron fields in  ${\cal N} =1$ SUSY QED there are two 
contributions to the electron self-energy corresponding to the graph depicted 
in Fig.~\ref{fig-electron}b. The first one corresponding the left selectron field  equals 
\be 
-i {_{(bL)}
\Sigma} (k) = (-ie \sqrt{2})^2 \frac{1}{2}
 \int \frac{d^4p}{(2\pi )^4}
\big[ i \tilde S^+(p+k) P_L i \tilde D^{\rm sym} (p) P_R
+  i\tilde S^{\rm sym} (p)  P_L i\tilde D^-(p-k) P_R \big],
\ee
which is 
\be
\label{Si-k-b-1}
{_{(bL)} \Sigma} (k) = i  e^2
 \int \frac{d^4p}{(2\pi )^4}
\big[ \tilde S^+(p+k) P_L  \tilde D^{\rm sym} (p) P_R
+  \tilde S^{\rm sym} (p)  P_L \tilde D^-(p-k) P_R
\big].
\ee
Substituting the functions $\tilde D^{\pm}$, $\tilde D^{\rm sym}$ and
$\tilde S^{\pm}$, $\tilde S^{\rm sym}$ given by
Eqs.~(\ref{G-pm}, \ref{G-sym}, \ref{Del-pm}, \ref{Del-sym})
into Eq.~(\ref{Si-k-b-1}), one finds in the Hard Loop Approximation
the following result
\ba
\label{Si-k-b-3}
_{(bL)} \Sigma (k) &=& e^2  \int \frac{d^3p}{(2\pi )^3} \,
 \frac{ f_{\tilde \gamma} ({\bf p}) +  f_s ({\bf p})}{E_p}  \,
\frac{P_L  p\sla P_R}{k\cdot p + i 0^+} ,
\ea
where we assumed that $f_s({\bf p}) = \bar f_s({\bf p})$.

Computing the contribution  corresponding to the graph depicted
in Fig.~\ref{fig-electron}b with the right selectron field, we get
\ba
\label{Si-k-b-3-R}
_{(bR)} \Sigma (k) &=& e^2  \int \frac{d^3p}{(2\pi )^3} \,
 \frac{ f_{\tilde \gamma} ({\bf p}) +  f_s ({\bf p})}{E_p}  \,
\frac{P_R p\sla P_L}{k\cdot p + i 0^+} .
\ea
Because $P_L  p\sla P_R + P_R p\sla P_L = p\sla$, the total contribution
given by the graph from Fig.~\ref{fig-electron}b equals
\ba
\label{Si-k-b-total}
_{(b)} \Sigma (k) &=&  e^2  \int \frac{d^3p}{(2\pi )^3} \,
 \frac{ f_{\tilde \gamma} ({\bf p}) +  f_s ({\bf p})}{E_p}  \,
\frac{ p\sla }{k\cdot p + i 0^+} .
\ea

\subsubsection{Final result}

The sum of expressions (\ref{Si-k-a-4}) and (\ref{Si-k-b-total}) gives the complete
electron self-energy
\ba
\label{Si-k-final}
\Sigma (k) &=& e^2  \int \frac{d^3p}{(2\pi )^3} \;
 \frac{ f_\gamma ({\bf p}) +  f_e ({\bf p}) + f_{\tilde \gamma} ({\bf p})  +  f_s ({\bf p})}{E_p}  \,
\frac{ p\sla }{k\cdot p + i 0^+} .
\ea
As seen, the electron self-energy has the same structure for the supersymmetric plasma 
and for its non-supersymmetric counterpart. 

\begin{figure}[t]
\centering
\includegraphics*[width=0.17\textwidth]{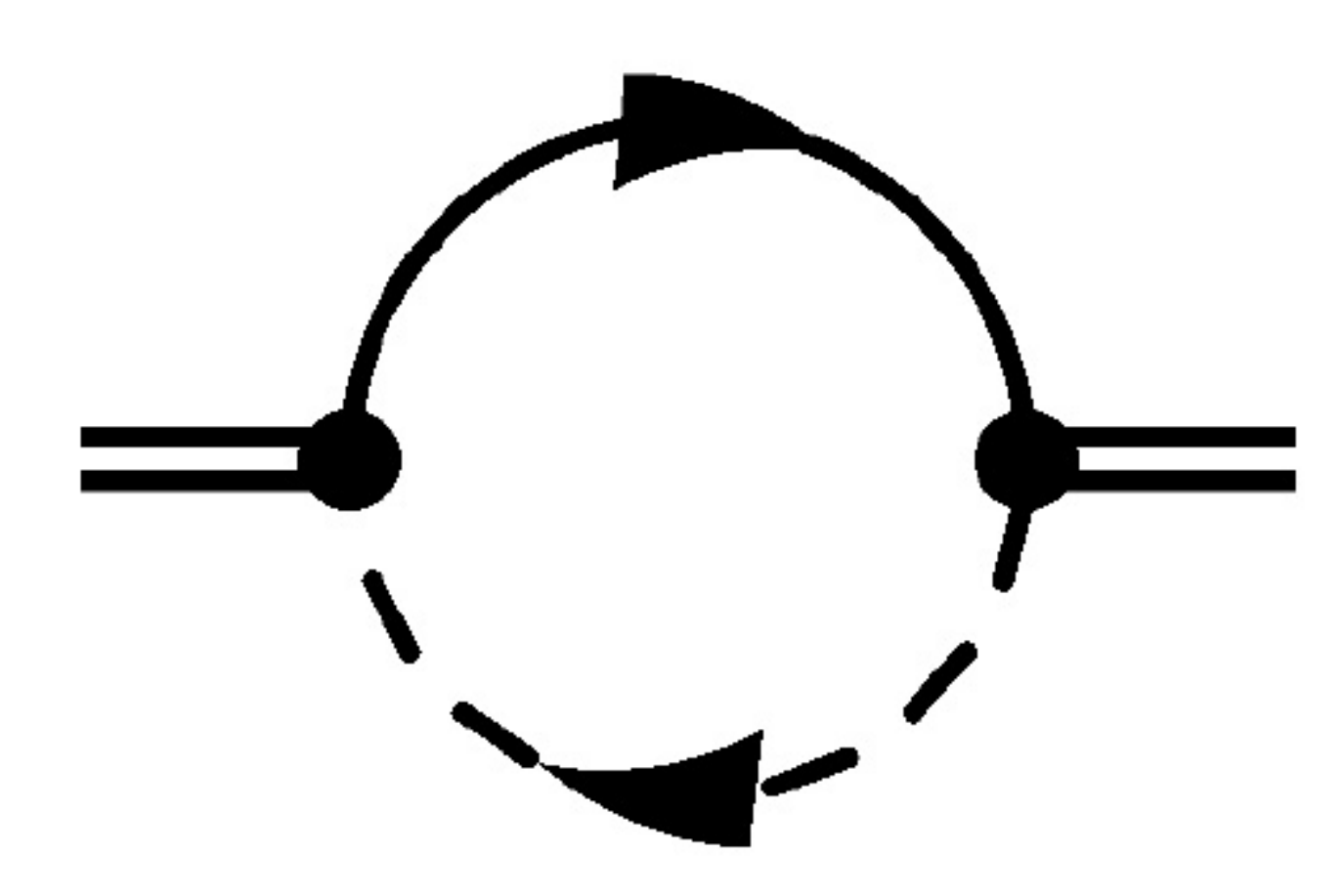}
\caption{Contribution to the photino self-energy.}
\label{fig-photino}
\end{figure}

\subsection{Photino self-energy}

The photino self-energy $\tilde \Pi$ can be defined by means of the Dyson-Schwinger 
equation 
\be 
i\tilde{\cal D} (k) = i \tilde D(k) 
+ i \tilde D(k) \, \big(- i\tilde \Pi(k) \big) \, i\tilde{\cal D}(k) ,
\ee 
where $\tilde {\cal D}$ and $\tilde D$ is the interacting and free photino propagator, 
respectively. The lowest order contribution to $\tilde \Pi$ is given by the diagram shown 
in Fig.~\ref{fig-photino}.  The solid, dashed and double-solid lines denote, respectively, 
electron, selectron and photino fields.  Since there are two selectron fields in 
${\cal N} = 1 $ SUSY QED there are two contributions represented by the diagram 
corresponding to the left and right selectrons. Appropriately modifying  
Eq.~(\ref{Pi-k-e-1}), one gets
\be 
\label{Pi-k-f-1} 
-i {_{(L)}} \tilde \Pi(k) =
\frac{1}{2}(-ie\sqrt{2})^2 \int \frac{d^4p}{(2\pi )^4} \big[P_R
iS^+(p+k)P_L i \tilde S^{\rm sym}(p) + P_R i S^{\rm sym}(p) P_L i
\tilde S^-(p-k) \big], 
\ee
where the contribution from left selectrons is taken into account. Eq.~(\ref{Pi-k-f-1})
is trivially manipulated to
\be 
\label{Pi-k-f-2}  
{_{(L)}} \tilde \Pi(k) = ie^2 \int \frac{d^4p}{(2\pi )^4} 
\big[P_R S^+(p+k)P_L \tilde S^{\rm sym}(p)
+ P_R S^{\rm sym}(p) P_L \tilde S^-(p-k) \big]. 
\ee

Now one substitutes the functions $S^{\pm}$, $S^{\rm sym}$ and $\tilde S^{\pm}$, 
$\tilde S^{\rm sym}$ given by Eqs.~(\ref{S-pm}, \ref{S-sym}, \ref{Del-pm}, \ref{Del-sym}) 
into Eq.~(\ref{Pi-k-f-2}). Performing the integration over $p_0$ and changing ${\bf p}$ 
into $-{\bf p}$ in the terms representing antiparticles, we obtain
\ba 
\label{Pi-k-f-4} 
 {_{(L)}} \tilde \Pi(k)
&=& \frac{1}{2}e^2 \int \frac{d^3p}{(2\pi )^3E_p} 
\\ \nn  
&& \times \bigg\{
\Big[\frac{P_R
(p\sla + k\sla) P_L } {(p+k)^2 + i\, {\rm
sgn}\big((p+k)_0\big)0^+}-\frac{P_R (p\sla -k\sla) P_L} {(p-k)^2 -
i\, {\rm sgn}\big((p-k)_0\big)0^+}\Big] \big( 2f_s({\bf p})+1\big)
\\ \nn 
&& + \Big[\frac{P_R p\sla P_L } {(p+k)^2 +
i\, {\rm sgn}\big((p+k)_0\big)0^+} - \frac{P_R p\sla P_L} {(p-k)^2
- i\, {\rm sgn}\big((p-k)_0\big)0^+}\Big] \big( 2f_e({\bf p}) - 1\big)
\bigg\},
\ea
where we assumed that $f_s({\bf p}) = \bar f_s({\bf p})$ and 
$f_e({\bf p}) = \bar f_e({\bf p})$. Adopting the Hard Loop Approximation, one gets
\ba 
\label{Pi-k-f-6}  
{_{(L)}} \tilde \Pi(k)&=& e^2
\int \frac{d^3p}{(2\pi )^3} \; \frac{f_s({\bf p})+f_e({\bf p})}{E_p} \;
 \frac{P_R p\sla P_L }{k\cdot p + i 0^+} . 
\ea

Since the contribution to the photino self-energy coming from right selectrons,
which is obtained in the same way, reads 
\be 
\label{Pi-k-f-7}  
{_{(R)}} \tilde \Pi(k) = e^2
\int \frac{d^3p}{(2\pi )^3} \; \frac{f_s({\bf p})+f_e({\bf p})}{E_p} \;
 \frac{P_L p\sla P_R }{k\cdot p + i 0^+} ,
\ee
one finds, using the well-known identity $P_R p\sla P_L + P_L p\sla P_R = p\sla$, 
the complete photino self-energy as
\be 
\label{Pi-k-f-final}
\tilde \Pi(k) = e^2 \int \frac{d^3p}{(2\pi )^3} \; 
\frac{f_s({\bf p})+f_e({\bf p})}{E_p} \; \frac{ p\sla }{k\cdot p + i 0^+} .
\ee
As seen, the photino self-energy (\ref{Pi-k-f-final}) has the same structure as 
the electron self-energy  (\ref{Si-k-final}).

\subsection{Selectron self-energy}

\begin{figure}[t]
\centering
\includegraphics*[width=0.9\textwidth]{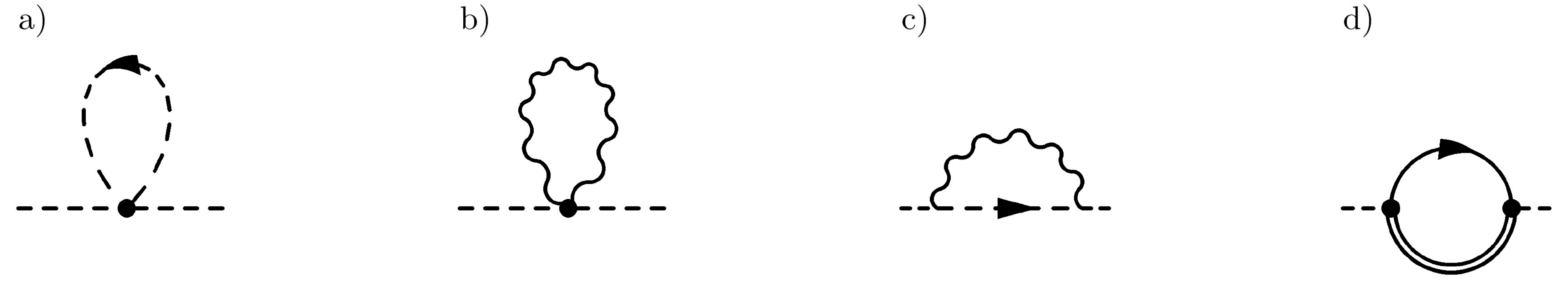}
\caption{Contributions to the selectron self-energy.}
\label{fig-selectron}
\end{figure}

The selectron self-energy $\tilde \Sigma$ can be defined by means
of the Dyson-Schwinger equation 
\be 
i\tilde{\cal S} (k) = i \tilde S (k) 
+ i \tilde S(k) \, i\tilde \Sigma (k)  \, i\tilde{\cal S}(k) , 
\ee 
where $\tilde{\cal S}$ and $\tilde S$ is the interacting and free propagator, 
respectively. The lowest order contributions to $\tilde \Sigma$ are given by 
four diagrams shown in Fig.~\ref{fig-selectron}. The solid, wavy, dashed and 
double-solid lines denote, respectively, electron, photon, selectron and photino 
fields. Below we compute the self-energy of left selectron. The result for right 
selectron is the same.

\subsubsection{Selectron tadpole}

There are two contributions represented by the graph depicted in Fig.~\ref{fig-selectron}a, 
as the tadpole  line corresponds to either left or right selectron. In the first case we have
\be 
\label{Si-k-s-t-1} 
i {_{(aL)}}\tilde \Sigma_L(k) = -2i e^2  \int \frac{d^4p}{(2\pi )^4}  i\tilde S^<(p) .
\ee
Substituting the function $\tilde S^<$ given by Eq.~(\ref{Del-<}) into Eq.~(\ref{Si-k-s-t-1}), 
one finds
\be 
\label{Si-k-s-t-2}
{_{(aL)}}\tilde \Sigma_L(k) = - e^2  \int \frac{d^3p}{(2\pi )^3}  \,
\frac{2f_s({\bf p})+1}{E_p}  , 
\ee
where the equality $\bar f_s({\bf p}) = f_s({\bf p})$ was assumed. The second contribution
corresponding to the right-selectron field equals
\be 
\label{Si-k-s-t-3} 
i {_{(aR)}}\tilde \Sigma_L (k) = i e^2
\int \frac{d^4p}{(2\pi )^4}  i\tilde S^<(p) ,
\ee
and it gives
\be 
\label{Si-k-s-t-4} 
_{(aR)}\tilde \Sigma_L (k) = \frac{1}{2}e^2 \int \frac{d^3p}{(2\pi )^3}  \,
\frac{2f_s({\bf p})+1}{E_p}  . 
\ee
Summing up the contributions (\ref{Si-k-s-t-2}, \ref{Si-k-s-t-3}), one
finds the following complete result of the  selectron tadpole
\be 
\label{Si-k-s-t-5} 
_{(a)}\tilde \Sigma_L (k) =   - \frac{1}{2}e^2  \int \frac{d^3p}{(2\pi )^3}  \,
\frac{2f_s({\bf p}) +1}{E_p}  . 
\ee

\subsubsection{Photon tadpole}

The contribution to the selectron self-energy coming from the photon tadpole shown 
in Fig.~\ref{fig-selectron}b equals
\be 
\label{Si-k-s-d-1} 
i {_{(b)}}\tilde \Sigma_L (k) = i e^2 g^{\mu \nu}
\int \frac{d^4p}{(2\pi )^4}  i D^<_{\mu \nu}(p) ,
\ee
where the symmetry factor $1/2$ is included. Eq.~(\ref{Si-k-s-d-1}) gives 
\be 
\label{Si-k-s-d-2} 
_{(b)}\tilde \Sigma_L (k) =  - 2 e^2
 \int \frac{d^3p}{(2\pi )^3}  \,
\frac{2f_\gamma({\bf p})+1}{E_p} ,
\ee
when the function $D^<_{\mu \nu}$ (\ref{D-<}) is substituted into Eq.~(\ref{Si-k-s-d-1}).

\subsubsection{Selectron-photon loop}

The contribution represented by the graph depicted in Fig.~\ref{fig-selectron}c equals
\be  
\label{Si-k-s-p-1} 
i {_{(c)}}\tilde \Sigma_L (k) =
\frac{1}{2} (-i e)^2  \int \frac{d^4p}{(2\pi )^4} 
\big[(p+2k)^\mu i \tilde S^+_{\mu \nu}(p+k) \, (p+2k)^\nu i D^{\rm sym}(p) 
+ (p+k)^\mu i \tilde S^{\rm sym}_{\mu \nu}(p) \, (p+k)^\nu i D^-(p-k) \big] ,
\ee
which after the substitution of the functions $D^{\pm}_{\mu \nu}, \, D^{\rm sym}_{\mu \nu}$ 
and $\tilde S^{\pm}, \, \tilde S^{\rm sym}$ in the form (\ref{D-pm}, \ref{D-sym}, \ref{Del-pm},
\ref{Del-sym}) leads to
\ba  
\label{Si-k-s-p-3} 
{_{(c)}}\tilde \Sigma_L(k) & = & \frac{1}{4} e^2 \int \frac{d^3p}{(2\pi )^3 E_p} 
\\ \nn
&& \times 
\bigg[ \bigg(\frac{(p+2k)^2 } {(p+k)^2 + i\, {\rm sgn}\big((p+k)_0\big)0^+} 
       + \frac{(p-2k)^2 } {(p-k)^2 - i\, {\rm sgn}\big((p-k)_0\big)0^+} \bigg)
\big(2f_s({\bf p}) +1\big) 
\\ \nn 
&&+ \bigg(\frac{(p+k)^2} {(p-k)^2 - i\, {\rm sgn} \big((p-k)_0\big)0^+}
+\frac{(p-k)^2} {(p+k)^2 + i\, {\rm sgn}\big((p+k)_0\big)0^+}\bigg) 
\big(2 f_\gamma({\bf p}) + 1\big)
\bigg] , 
\ea
where we assumed that $\bar f_s({\bf p}) = f_s({\bf p})$.  Within the Hard Loop
Approximation, one obtains
\ba 
\label{Si-k-s-p-4} 
{_{(c)}}\tilde \Sigma_L (k) 
& = &  \frac{e^2}{2} \int \frac{d^3p}{(2\pi )^3}
 \frac{4f_\gamma({\bf p}) -2f_s({\bf p})+1}{E_p} .
\ea
We note that the sum of the contributions (\ref{Si-k-s-d-2}, \ref{Si-k-s-p-4}),  which equals
\ba 
\label{Si-scalar-QED} 
{_{(b+c)}}\tilde \Sigma_L (k) 
& = & - \frac{e^2}{2} \int \frac{d^3p}{(2\pi )^3}
 \frac{4f_\gamma({\bf p}) +2f_s({\bf p})+3}{E_p} ,
\ea
represents the scalar self-energy of scalar QED which for equilibrium plasma
was discussed in {\it e.g.} \cite{Kraemmer:1994az} within the  imaginary-time 
formalism.

\subsubsection{Electron-photino loop}

The graph depicted in Fig.~\ref{fig-selectron}d provides
\be
\label{Si-k-r-1} 
i _{(d)} \tilde \Sigma_L (k) = (-1)(-ie \sqrt{2})^2 \frac{1}{2}
 \int \frac{d^4p}{(2\pi )^4}
{\rm Tr}\big[ P_R i S^+(p+k) P_L i \tilde D^{\rm sym} (p) + P_R
iS^{\rm sym} (p)  P_L i \tilde D^-(p-k)  \big].
\ee
Substituting the functions $\tilde D^{\pm}$, $\tilde D^{\rm sym}$ and $ S^{\pm}$, 
$ S^{\rm sym}$ given by Eqs.~(\ref{S-pm}, \ref{S-sym}, \ref{Del-pm}, \ref{Del-sym})
into Eq.~(\ref{Si-k-r-1}) and repeating the same steps which were made in the 
previous subsections, we find in the Hard Loop Approximation the following
expression
\be 
\label{Si-k-r-3} 
_{(d)} \tilde \Sigma_L (k) =  -2 e^2  \int \frac{d^3p}{(2\pi )^3} \,
\frac{ f_{\tilde \gamma} ({\bf p}) +  f_e ({\bf p})-1}{E_p}, 
\ee
where we assumed that $f_e({\bf p}) = \bar f_e({\bf p})$.

\subsubsection{Final result}

The sum of contributions (\ref{Si-k-s-t-5}, \ref{Si-k-s-d-2}, \ref{Si-k-s-p-4}) and (\ref{Si-k-r-3})
gives the complete self-energy of left selectron 
\be
\label{Si-k-r-final}  
\tilde \Sigma (k) = -2 e^2 \int \frac{d^3p}{(2\pi )^3} \,
\frac{f_e ({\bf p}) + f_\gamma ({\bf p}) +  f_s ({\bf p})+  f_{\tilde \gamma} ({\bf p}) }{E_p}, 
\ee
which equals the complete self-energy of right selectron. For this reason the index $L$ 
is dropped. As seen, the self-energy (\ref{Si-k-r-final}) is independent of $k$ and because 
of supersymmetry it vanishes in the vacuum limit when all the distribution functions 
are zero. This is also effect of the supersymmetry that the distribution functions of electrons
and of selectrons enter  the formula (\ref{Si-k-r-final}) with the coefficients equal to
each other. The same is true for the distribution functions of photons and of photinos.

\section{Effective Action}
\label{sec-eff-action} 

The Hard Loop approach can be formulated in an elegant and compact way by 
using the effective action which was first derived for equilibrium plasmas in
\cite{Taylor:1990ia,Frenkel:1991ts,Braaten:1991gm} within the thermal field 
theory. It was also rederived in terms of quasiclassical kinetic theory 
\cite{Blaizot:1993be,Kelly:1994dh}. Later on a generalization of the action to  
anisotropic systems was given in \cite{Pisarski:1997cp,Mrowczynski:2004kv}. 

The form of self-energy constrains the possible structure of the respective effective action. 
Since the self-energy of a given field is the second functional derivative of the action 
with respect to the field, one writes
\ba
\label{action-A-1}
{\cal L}^{(A)}_2(x) &=&  
\frac{1}{2} \int d^4y \; A_\mu(x) \Pi^{\mu \nu}(x-y) A_\nu(y) , 
\\ [2mm] 
\label{action-Psi-1}
{\cal L}^{(\Psi)}_2(x) &=& 
\int d^4y \; \bar{\Psi}(x) \Sigma (x-y) \Psi (y) ,
\\ [2mm] 
\label{action-Lambda-1}
{\cal L}^{(\Lambda)}_2(x) &=&  
 \frac{1}{2}
\int d^4y \; \bar\Lambda (x) \tilde\Pi(x-y) \Lambda (y) , 
\\ [2mm] 
\label{action-Phi-1}
{\cal L}^{(\phi_{L,R})}_2(x) &=&  
\int d^4y \; \phi_{L,R}^*(x) \tilde\Sigma(x-y) \phi_{L,R} (y) , 
\ea
where the self-energies are given by the formulas  
(\ref{Pi-k-final}, \ref{Si-k-final}, \ref{Pi-k-f-final}, \ref{Si-k-r-final}), respectively.
The subscript `2' indicates that the effective actions above only generate 
two-point functions. To generate $n$-point functions these actions 
need to be extended to a gauge invariant form. In the Abelian gauge theory 
studied here, the actions (\ref{action-A-1}, \ref{action-Lambda-1} ,\ref{action-Phi-1})
are gauge invariant and complete. The action (\ref{action-Psi-1}) requires a simple 
modification - the ordinary derivative is replaced by the covariant one in the
final expression. Repeating the calculations described in detail in 
\cite{Mrowczynski:2004kv}, one finds the Hard Loop effective action
of ${\cal N}=1$ SUSY QED as
\ba
{\cal L}_{\rm HL} 
&=& 
-\frac{1}{4}F^{\mu \nu} F_{\mu \nu} +  i\bar \Psi D\!\sla \Psi
+\frac{i}{2} \bar \Lambda \partial \sla \Lambda
+(D_\mu \phi_L)^*(D^\mu \phi_L) + (D_\mu^* \phi_R)(D^\mu \phi_R^*)
\\ \nn 
&& + \; {\cal L}^{(A)}_{\rm HL} +{\cal L}^{(\Psi)}_{\rm HL}
+ {\cal L}^{(\Lambda)}_{\rm HL} + {\cal L}^{(\phi_L)}_{\rm HL}
+{\cal L}^{(\phi_R)}_{\rm HL} ,
\ea
where
\ba
\label{action-A-2}
{\cal L}^{(A)}_{\rm HL} &=&  
4e^2  \int \frac{d^3p}{(2\pi )^3} \,
\frac{f_e({\bf p})+f_s({\bf p})}{E_p} \,
F_{\mu \nu} (x) {p^\nu p^\rho \over (p \cdot \partial)^2} F_\rho^{\;\;\mu} (x) , 
\\ [2mm] 
\label{action-Psi-2}
{\cal L}^{(\Psi)}_{\rm HL} &=&   i e^2  
\int \frac{d^3p}{(2\pi )^3} \; \frac{ f_\gamma ({\bf p}) +  f_e ({\bf p}) 
+ f_{\tilde \gamma} ({\bf p})  +  f_s ({\bf p})}{E_p}  \,
\bar{\Psi}(x) {p \cdot \gamma \over p\cdot D} \Psi (x) ,
\\ [2mm] 
\label{action-Lambda-2}
{\cal L}^{(\Lambda)}_{\rm HL} &=&  
i e^2 \int \frac{d^3p}{(2\pi )^3} \; 
\frac{f_s({\bf p})+f_e({\bf p})}{E_p} \; 
\bar\Lambda (x) {p \cdot \gamma \over p\cdot \partial} \Lambda (y) , 
\\ [2mm] 
\label{action-Phi-2}
{\cal L}^{(\phi_{L,R})}_{\rm HL} &=&  -
2 e^2 \int \frac{d^3p}{(2\pi )^3} \,
\frac{f_e ({\bf p}) + f_\gamma ({\bf p}) +  f_s ({\bf p}) +  f_{\tilde \gamma} ({\bf p}) }{E_p}
\; \phi_{L,R}^*(x) \phi_{L,R} (x) .
\ea

The actions  (\ref{action-A-1}, \ref{action-Psi-1}, \ref{action-Lambda-1}, \ref{action-Phi-1})
are obtained from the self-energies but the reasoning can be turned around. As argued 
in \cite{Frenkel:1991ts,Braaten:1991gm}, the actions of gauge bosons (\ref{action-A-2}), 
charged fermions (\ref{action-Psi-2}) and charged scalars (\ref{action-Phi-2}) are of
unique gauge invariant form.  Therefore, the respective self-energies can be, in principle, 
inferred from the known QED self-energies with some help from supersymmetry arguments. 
In the case of photino self-energy, which is of our main interest, the explicit computation 
seems to be unavoidable.

\section{Collective modes and conclusion}
\label{sec-modes}

When the self-energies computed in Sec.~\ref{sec-self-energies} are substituted into 
the dispersion equations presented in Sec.~\ref{sec-dis-eqs}, collective modes 
can be found as solutions of the equations. Below we briefly discuss the  
photon, electron, photino and selectron excitations.

\begin{itemize}

\item
The structure of polarization tensor (\ref{Pi-k-s-total}) is such as that in the usual 
non-supersymmetric QED plasma. It also coincides with the gluon polarization
tensor of QCD plasma. Therefore, the spectrum of collective excitations 
of gauge bosons is in all three cases the same. In equilibrium plasma we have 
the longitudinal (plasmon) mode and the transverse one which are discussed 
in {\it e.g.} the textbook \cite{lebellac}. When the plasma is out of equilibrium 
there is a whole variety of possible collective excitations. In particular, there 
are unstable modes, see  {\it e.g.} the review  \cite{Mrowczynski:2007hb}, which 
exponentially grow in time and strongly influence the system's dynamics. 

\item
The form of electron self-energy (\ref{Si-k-final}) happens to be the same as 
in the usual non-supersymmetric QED plasma. The quark self-energy in QCD plasma
has the same form. Therefore, we have identical spectrum of excitations of 
charged fermions in the three systems. In equilibrium plasma there 
two modes, see in {\it e.g.} the textbook \cite{lebellac}, of opposite helicity 
over chirality ratio.  One mode corresponds to the positive energy fermion, 
another one, sometimes called a plasmino, is a specific medium effect.  In 
non-equilibrium plasma the spectrum of fermion collective excitations changes
but no unstable modes have been found even for an extremely  anisotropic 
momentum distribution  \cite{Mrowczynski:2001az,Schenke:2006fz}.

\item
The photino self-energy (\ref{Pi-k-f-final}) has a structure identical to the electron 
self-energy (\ref{Si-k-final}) and thus the spectra of collective excitations 
are also identical.  When the plasma momentum distribution is anisotropic
and unstable photon modes occur, the photino modes remain stable.
Supersymmetry  does not change anything here. 

\item
The selectron self-energy (\ref{Si-k-r-final}) is  independent of momentum,
it is negative and real. Therefore,  $\tilde \Sigma $ can be written as 
$\tilde \Sigma = - m^2_{\rm eff}$ where $m_{\rm eff}$ is the 
effective selectron mass. Then, the solutions of dispersion equation
(\ref{dis-eq-selectron}) are $E_p = \pm \sqrt{m^2_{\rm eff} + {\bf p}^2}$.

\end{itemize}

We conclude our considerations by saying that the collective modes in
ultrarelativistic plasma of  ${\cal N} =1$ SUSY QED are essentially the same
as in ultrarelativistic electromagnetic plasma of electrons, positrons and  
photons. 

\section*{Acknowledgments}

We are very grateful to Margaret Carrington  for helpful correspondence.
This work was partially supported by the Polish Ministry of Science  and Higher  
Education under grants N~N202~204638 and 667/N-CERN/2010/0.

\appendix*

\section{Green's functions of Keldysh-Schwinger formalism}

We present here the retarded, advanced and unordered Green's functions which 
are usually labeled with the indices $+, -, >, <$, respectively. The form of these 
functions for free non-equilibrium fields can be found in {\it e.g.} \cite{Mrowczynski:1992hq}.

\subsection{Photons}

The functions of interest for the free electromagnetic field in the Feynman gauge are
\ba
\label{D-pm}
D^{\pm}_{\mu \nu}(p) &=& -  \frac{g_{\mu \nu}}{p^2\pm i\, {\rm sgn}(p_0)0^+},
\\
\label{D->}
D^>_{\mu \nu}(p) &=& \frac{i\pi g_{\mu \nu}}{E_p}
\Big(\delta (E_p - p_0) \big[ f_\gamma ({\bf p}) +1\big]
+ \delta (E_p + p_0) f_\gamma (-{\bf p}) \Big),
\\
\label{D-<}
D^<_{\mu \nu}(p) &=& \frac{i\pi  g_{\mu \nu}}{E_p}
\Big( \delta (E_p - p_0) f_\gamma ({\bf p})
+ \delta (E_p + p_0)  \big[ f_\gamma (-{\bf p}) + 1\big] \Big),
\\
\label{D-sym}
D^{\rm sym}_{\mu \nu}(p) &\equiv& D^>_{\mu \nu}(p) + D^<_{\mu \nu}(p)
= \frac{i\pi g_{\mu \nu}}{E_p}
\Big( \delta (E_p - p_0) \big[2 f_\gamma({\bf p}) +1\big]
+ \delta (E_p + p_0) \big[2 f_\gamma(-{\bf p})+ 1\big]\Big),
\ea
where $f_\gamma({\bf p})$ is the distribution function of photons which are assumed
to be unpolarized. The function is normalized in such a way that the photon density
is given as
\be
n_\gamma = 2 \int \frac{d^3p}{(2\pi)^3}\, f_\gamma ({\bf p}) ,
\ee
where the factor of 2 takes into account two photon spin states.

One checks that the functions (\ref{D-pm}, \ref{D->}, \ref{D-<}) obey
the required identity
\be
\label{id-D}
D^>_{\mu \nu}(p) - D^<_{\mu \nu}(p) = D^+_{\mu \nu}(p) - D^-_{\mu \nu}(p) .
\ee
The left-hand side of Eq.~(\ref{id-D}) equals
\be
\label{D>-D<}
D^>_{\mu \nu}(p) - D^<_{\mu \nu}(p) = \frac{i\pi g_{\mu \nu}}{E_p}
\big(\delta (E_p - p_0)  - \delta (E_p + p_0)  \big) =
2i\pi \, g_{\mu \nu} \delta (p^2) \big(\Theta(p_0)  - \Theta(- p_0) \big).
\ee
Using the well-known relation
\be
\frac{1}{x \pm i0^+} = {\cal P}\frac{1}{x } \mp i \pi \delta(x),
\ee
one immediately shows that the right-hand side of Eq.~(\ref{id-D}) equals
the expression (\ref{D>-D<}).

\subsection{Electrons}

The functions for the free massless electron field are
\ba
\label{S-pm}
S^{\pm}(p) &=&  \frac{p\sla}{p^2\pm i\, {\rm sgn}(p_0)0^+},
\\
\label{S->}
S^>(p) &=& \frac{i\pi}{E_p} p\sla \Big( \delta (E_p - p_0)  \big[ f_e({\bf p}) -1\big]
+ \delta (E_p + p_0) \bar f_e(-{\bf p}) \Big),
\\
\label{S-<}
S^<(p) &=& \frac{i\pi}{E_p} p\sla \Big( \delta (E_p - p_0)  f_e({\bf p})
+ \delta (E_p + p_0) \big[ \bar f_e(-{\bf p}) - 1\big] \Big),
\\
\label{S-sym}
S^{\rm sym}(p) &\equiv& S^>(p) + S^<(p)
= \frac{i\pi}{E_p} p\sla \Big( \delta (E_p - p_0)   \big[2 f_e({\bf p}) -1\big]
+  \delta (E_p + p_0)  \big[2 \bar f_e(-{\bf p})- 1\big] \Big),
\ea
where $f_e({\bf p})$ and $\bar f_e({\bf p})$ are the distribution functions of electrons
and of positrons, respectively.  We assume here that both electrons and positrons 
are unpolarized. The distribution functions are normalized in such a way that the electron 
density equals
\be
n_e = 2 \int \frac{d^3p}{(2\pi)^3}\, f_e({\bf p}) ,
\ee
where the factor of 2 takes into account two spin states of each electron.
The functions (\ref{S-pm}, \ref{S->}, \ref{S-<}) are checked to obey 
the identity $S^>(p) - S^< (p) = S^+(p) - S^-(p)$.

\subsection{Photinos}

The functions for the free photino field read
\ba
\label{G-pm}
\tilde D^{\pm}(p) &=&  \frac{p\sla}{p^2\pm i\, {\rm sgn}(p_0)0^+},
\\
\tilde D^>(p) &=& \frac{i\pi}{E_p} p\sla \Big( \delta (E_p - p_0)
\big[ f_{\tilde \gamma} ({\bf p}) -1\big]
+ \delta (E_p + p_0)  f_{\tilde \gamma} (-{\bf p}) \Big),
\\
\tilde D^<(p) &=& \frac{i\pi}{E_p} p\sla \Big( \delta (E_p - p_0)
f_{\tilde \gamma}({\bf p}) + \delta (E_p + p_0) \big[ f_{\tilde
\gamma} (-{\bf p}) - 1\big] \Big),
\\
\label{G-sym}
\tilde D^{\rm sym}(p) &\equiv& \tilde D^>(p) + \tilde D^<(p)
= \frac{i\pi}{E_p} p\sla \Big( \delta (E_p - p_0)   \big[2 f_{\tilde \gamma} ({\bf p}) -1\big]
+  \delta (E_p + p_0)  \big[2 f_{\tilde \gamma}(-{\bf p}) - 1\big] \Big),
\ea
where $f_{\tilde \gamma}({\bf p})$ is the distribution function of photinos
which are assumed to be unpolarized. The function is normalized in
such a way that the photino density is given as
\be
n_{\tilde \gamma} = 2 \int \frac{d^3p}{(2\pi)^3}\, f_{\tilde \gamma}({\bf p}) ,
\ee
where the factor of 2 takes into account two photino spin states.
One checks that the required relation 
$\tilde D^>(p) - \tilde D^< (p) = \tilde D^+(p) - \tilde D^-(p) $
is satisfied.

\subsection{Selectrons}

The functions for the free selectron field are
\ba
\label{Del-pm}
\tilde S^{\pm}(p) &=&  \frac{1}{p^2\pm i\, {\rm sgn}(p_0)0^+},
\\
\label{Del->}
\tilde S^>(p) &=& -\frac{i\pi}{E_p} \Big( \delta (E_p - p_0) \big[
f_s({\bf p}) +1\big] +  \delta (E_p + p_0) \bar f_s(-{\bf p})
\Big) ,
\\
\label{Del-<}
 \tilde S^<(p) &=& - \frac{i\pi}{E_p} \Big( \delta
(E_p - p_0) f_s({\bf p}) +  \delta (E_p + p_0)  \big[ \bar
f_s(-{\bf p}) + 1\big] \Big) ,
\\
\label{Del-sym}
\tilde S^{\rm sym}(p) &\equiv& \tilde S^>(p) + \tilde S^<(p)
= - \frac{i\pi}{E_p}\Big( \delta (E_p - p_0) \big[2 f_s({\bf p}) +1\big]
+  \delta (E_p + p_0) \big[2 \bar f_s(-{\bf p})+ 1\big] \Big),
\ea
where $f_s({\bf p})$ is the distribution function of left or right selectrons 
and $\bar f_s({\bf p})$ is the distribution function of left or right  antiselectrons.
We assume that the distribution functions of left and right (anti-)selectrons 
are equal to each other. The functions (\ref{Del-pm}, \ref{Del->}, 
\ref{Del-<}) obey the identity 
$\tilde S^>(p) - \tilde S^< (p) = \tilde S^+(p) - \tilde S^-(p)$.


\end{document}